# The effect of color-coding on students' perception of learning in introductory mechanics

Brianna S. Dillon Thomas*, Scott Carr, and Siming Guo
*Department of Physics and Engineering Science, Coastal Carolina University*
*109 Chanticleer Dr. East, Conway, SC 29526*

## Abstract

We designed three color-coding schemes to identify related information across representations and to differentiate distinct information within a representation in slide-based instruction for calculus-based introductory mechanics. We found that students had generally favorable opinions on the use of color and that the few negative criticisms are easily addressed through minor modifications to implementation. Without having the color-coding schemes pointed out to them, a modest but consistent minority of students who found color helpful also described the color-coding schemes implemented, and about a quarter described the use of color in physics contexts as helpful even if they did not describe color-coding. We found that students particularly favored using color in mathematics and color-coding used to identify related variables, verbal definitions, and diagram elements. We additionally found that on average 40% of students found color to be helpful in matching and connecting related information or in separating and distinguishing distinct information, which were the motivating reasons for employing the color-coding schemes.

## I. Introduction and Motivation

Students in introductory physics courses often struggle with seeing algebraic expressions as physically meaningful and with the varied representations needed to model physical phenomena. The first struggle is a cognitive shift: in mathematics symbols are often interchangeable, while in physics they are standardized representations of physical quantities.[1] Mathematically proficient students still struggle to apply mathematics to quantify and interpret physical scenarios [2], and confusion about a variable's meaning is a factor, if not the most significant factor, in students' lower ability to correctly solve problems symbolically compared to solving them numerically [3-5]. The second struggle is developing a new skill. Students are presented with and expected to use an array of conceptually equivalent representations for physical phenomena: mathematics, verbal descriptions, graphs, sketches, and abstracted diagrams. Students are less adept at identifying related representations, recognizing which features of visual representations are relevant to a scenario, and connecting abstract representations to the physical world.[6-9]

To alleviate these challenges, one of us (BSDT) employed color-coding in introductory mechanics instructional materials. Color-coding implementations are characterized by using color strategically to convey information rather than for aesthetics only. We used color to identify related information across representations and differentiate distinct information within a representation.

Color's impact on learning and memory has been well explored in psychology, cognitive science, and education. Using colored instructional or presentation materials has long been known to positively influence test outcomes over black-and-white materials, even when color is simply for illustration, not coding information systematically.[10] Other studies on systematic color-coding among students of various ages have shown it improves the performance of students with weaker inherent ability to mentally organize and categorize salient visual information [11-12] and improves short and long-term retention of information.[13-14] A review of current psychology literature by Dzulkifli and Mustafar found that using color in instructional material improves memory by enhancing attention and increasing arousal (alertness to environment), and that high contrast, consistent color use in encoding material were key factors in improving memory performance. [15] An eye-tracking study by Ozcelik et al. supports these findings by showing that color-coding allows more efficient matching of text and images and more strongly directs attention to important information.[16]

*Corresponding author: bthomas1@coastal.edu    1

The benefits to color-coding instructional materials are supported broadly and suggest that color-coding physics instructional materials would help students with the previously described challenges. However, there is little systematic study of color-coding within physics education. Several studies of color-coding in mathematics education have mixed results, some indicating improvement in learning or in students' perceptions of learning and others showing no significant improvement but at least no harm.[17-19] The closest discipline-based study of color-coding found when starting this study was one by Reisslein, Johnson, and Reisslein in which color-coding introductory circuits materials was found to improve student learning outcomes and reduce perceived cognitive load.[20]

The merit of color-coding demonstrated in psychology and cognitive science combined with the relative scarcity of studying its application in discipline-specific contexts suggests that studying the impact of color-coding in physics instruction is worthwhile. The purpose of this study was to examine whether the use of color-coding verbal, mathematical, and visual representations in the instructional materials of an introductory mechanics class affected students' perceptions of their learning, and if so, what reasons they described for the perceived effect.

In Section II, we describe the color-coding intervention employed. In Section III we describe the study design and survey questions. In Section IV we describe the two sets of categories which arose from evaluating survey responses and were used to classify responses. In Section V we discuss the results of survey response categorization and observed trends regarding student perceptions of color and its utility in instructional materials, with conclusions and takeaways summarized in Section VI.

## II. Color-coding Intervention

For this study, we define color-coding as the purposeful or systematic use of color to identify related information across representations and to differentiate distinct information within a representation. We designed three color-coding schemes for use in slide-based instructional materials [21]:

- **Scheme 1: Equations and Definitions** – Same color used for a word in a verbal definition and the corresponding variable in the mathematical definition, with different colors used for each quantity (Fig 1).
- **Scheme 2: Equations and Diagrams** – Same color used for a variable in equation(s) and the corresponding portion of a diagram, with different colors used for each quantity. This was employed in both definitions (Fig 2a&b) and example solutions (Fig 2c).
- **Scheme 3: 2D Scenarios** – In two-dimensional scenarios for which vector decomposition was required, one color was used for x-components, another for y-components, and a third for anything shared between component equations, both in diagrams and in mathematical representations or worked-out example problems (Fig 3).

> Velocity is defined as the rate at which an object's position changes over time.
>
> $$\vec{v} = \frac{d\vec{r}}{dt} \quad \text{In 1D:} \quad v_x = \frac{dx}{dt}$$
>
> Subscript means "x-component of velocity"

**Figure 2.** (color online) *An example of color-coding Scheme 1 "Equations and Definitions", in which the same color is used in the verbal definition of a physical quantity and its corresponding variable in the mathematical definition Here, purple, red, blue, and yellow were used to code velocity, position, time, and rate of change respectively.*

(a) $\vec{p} = m\vec{v}$  $v = 18.3\ m/s$  $m = 1.2\ kg$

(b) $F_{AB}^k = \mu_k |F_{AB}^N|$  $F_{TB}^N$ Block  $F_{TB}^k$  $F_{EB}^G$  Table

(c) $\sum F = ma$

$$F_{TB}^C - F_{EB}^G - F_{HB}^C = m_B \cancel{a_B}^0$$

$$F_{TB}^C - m_B g - F_{HB}^C = 0N$$

$$F_{TB}^C = +m_B g + F_{HB}^C$$

$$F_{TB}^C = (3kg * 9.81 m/s^2) + (5N) = 34.43N$$

$F_{TB}^C$ ↑+x  $F_{EB}^G$  $F_{HB}^C = 5N$

**Figure 1.** (color online) *Examples of Scheme 2 "Equations and Diagrams", using the same color for the variable in an equation and the corresponding part of a diagram: (a, top left) defining momentum, (b, bottom left) defining kinetic friction, and (c, right) an example problem solution in which a person pushes downward on a book resting on a table. In each, the color used for the variable is also used for the color of the quantity's visual representation, e.g. in (a) mass "m" and the block with that mass are both colored blue while velocity "v" and the velocity arrow are colored red, in (b) the coefficient of kinetic friction, block, and table all colored shades of gold, and in (c) the gravitational force arrow in the free body diagram, and all variables and values used for calculating the gravitational force in the solution are colored purple.*

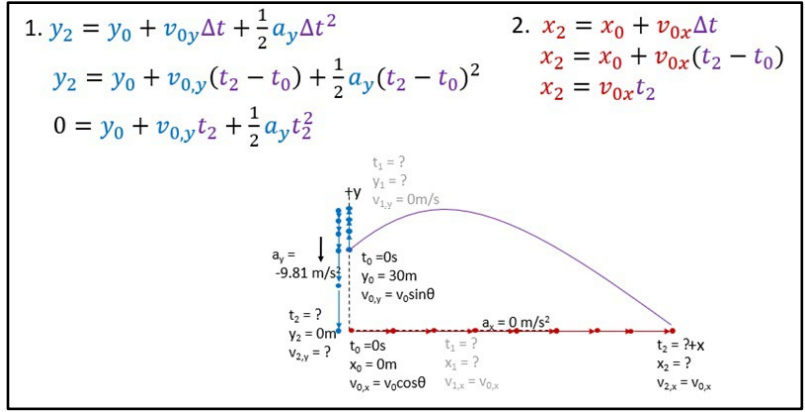

**Figure 3.** (color online) *An example of Scheme 3 "2D Scenarios", in which one color was used for x-components, another for y-components, and a third for anything shared between component equations, in both equations and diagrams. In this example, red and blue were used for x-component and y-component variables and diagram elements respectively, and purple was used for time in the equations and the 2D trajectory in the diagram.*

Schemes 1 and 2 were intended to facilitate identifying a symbolic variable's physical meaning by including a visual cue to connect the variable and its verbal name or diagrammatic representation, as suggested by similar applications in Ref 16. Scheme 3 was intended to address difficulties in decomposing a two-dimensional scenario into two one-dimensional problems, keeping components separate during operations, and knowing what to recombine. These challenges are comparable to those which motivated Schemes 1 and 2, and is supported by studies outside of physics which find that color-coding assists organizing and matching information.[11,12,16] Using Schemes 2 and 3 in example solutions was intended to highlight the connection between required set-up diagrams and the variables used (another example of decoding symbolic meaning) and to help students follow the logic of the symbolic algebra. This application is consistent with Madsen et al.'s suggestion in Ref. 7 to use visual cues to guide attention to salient features during problem-solving. When selecting colors for the schemes, using both red and green was avoided to account for the most common type of color-blindness, except for a few derivations which connected two related equations and had more than three or four distinct variables.

## III. Survey Design and Methodology

This study was implemented in a calculus-based introductory mechanics course at a predominantly undergraduate state university (enrollment ~10k students) in the southeastern United States, taught following Eric Mazur's *Principles and Practice of Physics* (1[st] ed) and using a flipped classroom format with recorded videos of voice-over slide presentations and worked example problems. The study had about 50 students in two one-semester cohorts, one each in Fall 2020 and Spring 2021.

The three color-coding schemes were employed in video lectures and example problems without mentioning their use or describing their purpose. Schemes 1 and 2 were used all semester. Scheme 3 was employed only in the last quarter of the semester because Mazur's topical schedule first covers all major mechanics topics exclusively in one dimension, then revisits each in two dimensions prior to rotational motion. Scheme 3 was employed primarily when introducing vector decomposition for each topic rather than on every 2D problem. Instructional materials were identical each semester, including narration.

Nine survey questions (Table I, columns 1-3) were administered throughout each semester, the first at the end of the first unit, and the last on the last day of class. Question 3 is omitted from the table and this report; it had a different focus and its responses had no bearing on our conclusions. Additional details regarding survey administration are in Appendix Section I.

The survey questions asked students what features of the course they found helpful, from an open-ended question (Question 1, 4, 9) to a progressively narrower focus on lesson materials (Q2) and the use of color in general (Q5) or in the context of the three schemes described in Section II (Q6-8) (see Table I columns 4-6). Since the color-coding schemes were never pointed out explicitly, this progression checked the extent to which students noticed the schemes without prompting. The most open-ended question was asked multiple times (Questions 1, 4, 9) to account for longer exposure to different schemes giving more opportunities to notice them. The remaining questions asked at most about the use of color in a particular *context* (Q6-8); no question asked about color-*coding* directly. These questions followed a pattern (see "Q5-8 Template" in Table I) which probed our research question: whether students considered the use of color and color-coding to have an impact on their learning ("what") and if so, what reasons they identified for that impact ("why").

**Table I.** *Survey questions numbered in order of appearance by week of semester and day of instruction. Columns 4 and 5 designate whether the question asked primarily about the use of color and whether it asked students to give a reason for what they found helpful or unhelpful. Col 6 identifies which schemes were used in class prior to the question's administration. Questions 5-8 followed a similar template; to aid seeing the differences between them the template is provided separately from the phrase unique to the question.*

| Question | Week and Day asked | Question Text | | Color Focus | Asks Reason | Schemes Used |
|---|---|---|---|---|---|---|
| 1 | W4, D11 | A) What is one thing about how this class is structured that is beneficial in helping you learn? B) What's one thing that could been better if _____ (fill in the blank with a concrete suggestion)? | | No | No | 1 & 2 |
| 2 | W6, D16 | Please comment on the format of [slides] in this class: quantity of info per slide, use of color and visuals, font size/type, amount of time per slide, ease of use as reference after class, etc etc. I'm curious to know what you find helpful and what you find a hindrance. | | No | No | 1 & 2 |
| 4 | W9, D26 | Same as Question 1 | | No | No | 1 & 2 |
| 5 | W10, D28 | *Q5-8 Template: Please comment on the use of color _________ this semester. What is helpful and why, what is not helpful/distracting and why, and what are you indifferent about?* | …in the [slides] and examples… | Yes, generally | Yes | All, but Scheme 3 only very recently |
| 6 | W11, D30 | | …in presenting equations and definitions… | Yes, Scheme 1 | Yes | All |
| 7 | W13, D36 | | ..in pairing equations or math with physics diagrams… | Yes, Scheme 2 | Yes | All |
| 8 | W14, D39 | | …in distinguishing 2D components… | Yes, Scheme 3 | Yes | All |
| 9 | W14, D40 | Same as Question 1 | | No | No | All |

## IV. Analysis

We used two classifications to categorize each survey response: a "Type" category for what feature was described as helpful or unhelpful, and a "Reason" category for the stated reason for the feature being helpful or unhelpful. First author BSDT evaluated survey responses for trends to define code categories prior to response coding with coauthors S.C. and S.G.

The "Type" categories designate both a response's opinion on the use of color (none, indifferent, unhelpful, helpful) and the extent to which a color-coding scheme was described in the response (not at all, part, whole but not systematic, or whole and systematic). Brief category descriptions are in Table II. For Questions 6, 7, and 8, which target specific schemes, the categories "Other color-coding" and "Other color-context" allow for differentiation between types of off-topic responses. Since the survey questions were open-ended, each response could have a mix of feedback with different parts categorized as different "Types".

The "Reason" categories describe in general terms what students did and did not find helpful about the use of color. While this study was motivated by specific student struggles, the "Reason" categories were not prescriptive, but rather emerged from evaluating survey responses for common themes. The themes that emerged pertained to the generic comprehension and attention benefits expected from psychology and cognitive science research, but also to three specific features related to common struggles and the motivations for employing color-coding. "Reason" categories are defined in Table III. Multiple "Reason" categories could apply to a single response if it made distinct statements that were not linked to each other. Each distinct statement was assigned a single category.

All three authors independently categorized survey responses according to "Type" and "Reason" categories. When aggregating results, "Type" and "Reason" categories were assigned based on consensus (at least two of the three agree on appropriateness of category). For a few percent of responses, there was no consensus on the appropriate "Type" and/or "Reason" category. These responses were assigned to the generic category with the most overlap with each authors' choice(s) ("Color generic" or "Uncategorizable" for "Type", and "Comprehension" or "Helpful, Other" for "Reason"). A thorough description of category assignment is in the Appendix Section II.C. All cases of disagreement between authors about categorization stemmed from the clarity of survey responses, not disagreement in understanding category definitions.

More detailed definitions of "Type" and "Reason" categories, example responses for each category, the nuances in determining appropriate categories for a response, and details about rate of agreement on category selection between authors are provided in Appendix, Section II.

**Table II.** *Brief definitions of the "Type" categories used to categorize survey responses.*

| Type Category | Opinion of color | Description of Scheme |
|---|---|---|
| **Color-coding** | Helpful | All parts of relevant scheme referenced AND acknowledges systematic use of color between parts |
| **Other color-coding** | Helpful | Same as "Color-coding", but for a scheme not referred to in question |
| **Color context** | Helpful | Only *part* of relevant scheme mentioned OR All parts of relevant scheme mentioned but systematic use of color between parts not acknowledged |
| **Other color context** | Helpful | Same as "Color context", but for a scheme not referred to in question |
| **Color generic** | Helpful | Does not identify color use in context of a scheme |
| **Color indifferent** | Neutral opinion | Yes or No both acceptable |
| **Color not helpful** | Not helpful | Yes or No both acceptable |
| **Other non-color** | None– color not mentioned | N/A |
| **Off-topic** | None – Did not address question | N/A |

**Table III.** *Definitions of the "Reason" categories used to categorize survey responses.*

| Reason Category | Description |
|---|---|
| *Specific Reasons* | |
| **Matching/connecting info** | Identifies either matching different representations of same physical entity (math/equation, visual/diagram, verbal/written), or connecting parts of different representations as being related. |
| **Separating/distinguishing** | Identifies being able to see different types/kinds/parts of information as being distinct from each other. Information may be visual, mathematical, verbal. |
| **Tracking** | Identifies being able to follow a piece of information through/across multiple sequential steps within the same representation. |
| *Generic Reasons* | |
| **Comprehension** | Identifies color as relates to learning generically, e.g. comprehension, understanding, visualizing, memory, and organizing information in the sense of "seeing related parts". |
| **Appearance** | Identifies color as relates to aesthetics, e.g. attention, interest, emphasis, importance, organization in the sense of "outlining/headers". |
| **Helpful, Other** | Describes color being helpful/positive in vague general terms only. |
| **Helpful, No Reason** | Does not indicate a reason for helpfulness. |
| **Not helpful, reason** | Indicates a reason for why color not liked or considered not helpful. Reasons noted but not coded separately. |

# V. Results and Discussion

Response rates for each "Type" category overall and for each question are in Fig 4. Response rates for each "Reason" category are in Fig 5, including subtotals for the proportion of responses which named one or more reasons specific to physics contexts ("Specific reason subtotal") or which named one or more general reasons ("Generic reason subtotal").

Among the identical open-ended questions, no responses to Questions 1 or 9 mentioned color; responses to those questions are excluded from totals. In response to Question 4, only one person described the use of color in course materials as generically beneficial for aesthetic reasons (categorized as Type "Color generic" and Reason "Appearance"). This response is included in the "All" totals of each figure.

## A. Overall perception of the use of color

For the remaining five questions, the use of color was generally perceived as helpful, with 75.3% of all responses identifying something positive about the question-relevant use of color (sum of "Color-coding", "Color context", and "Color generic" in Fig 4). Only 6.8% overall identified some aspect of color use as unhelpful, and only 2% expressed an exclusively negative opinion.

Favorable opinion of the use of color was substantial even when students were asked about the format of lesson slides broadly, with 47.6% of all responses describing color as helpful. Favorable opinions increased when students were specifically prompted to consider slides in relation to their use of color without reference to a particular context (Q5), and when prompted to consider a specific context (Q6-Q8). Question 8 (2D components) had the lowest rate of positive responses (69.4%) of all questions except Question 2 and the highest rate of negative or indifferent responses (19.4% combined). Question 7 (diagrams and equations) had the highest positive rate (91.9%) and lowest rate of negative or indifferent responses (5.4% combined).

The most common reason for considering color unhelpful was too many colors being confusing or distracting. This came up three times in Question 6 and twice in Question 8; due to the anonymity of responses there could be as few as three and as many as five participants who mentioned this issue. Another disliked colors being too similar. On three different questions, someone described red as a bad color choice compared to green or blue; it is reasonable to assume these were the same person.

Four responses described limitations to the context of color use. The most constructive criticism was that overuse was counter-productive: "It was helpful in the beginning to help show what each means but as time has progressed sometimes the use of color is severely overwhelming especially when it is equations we are familiar with and are only introducing a new way to use them" (Q7). Two others recognized color-coding: one found it confusing when color was used with no pattern, and one thought color-coding made it too easy to skip to the end of derivations.

Overall, the problems identified are either easily corrected or provide concrete suggestions for refining use of color-coding, such as avoiding it in complex scenarios for which there are more than a few distinct quantities or prioritizing its use when introducing novel information vs novel contexts of familiar information.

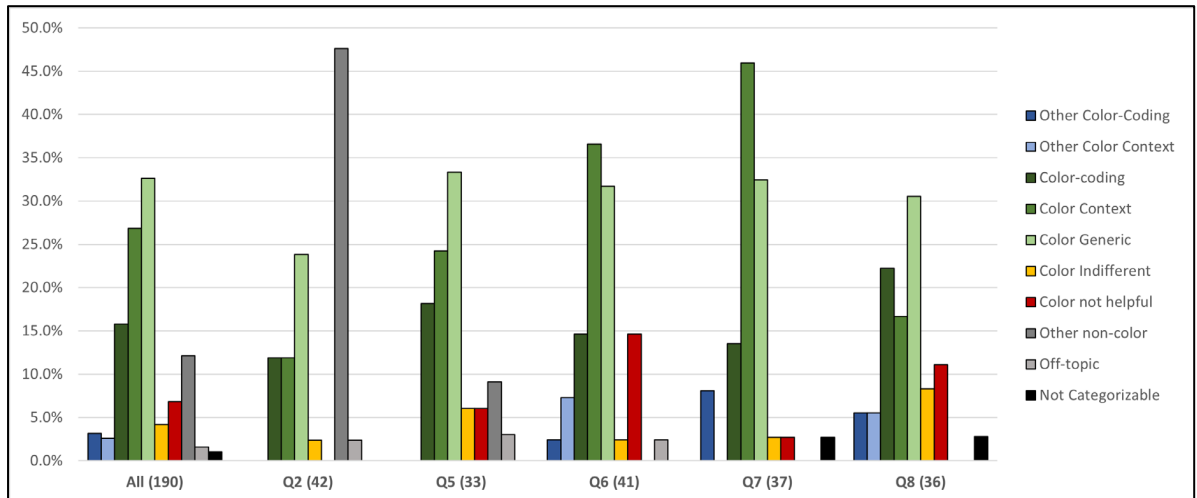

**Figure 4.** (color online) *Percent of responses within each "Type" category, for all responses collectively ("All") and for each question individually. The number in parenthesis after the cluster name is the number of responses to that question. Color-coding, color-context, color generic, other non-color, and off-topic are mutually exclusive with each other but not mutually exclusive with other categories, so percentages may not add up to 100%.*

*B. Strategic use of Color*

For each question, at least half of the responses which identified color as helpful also identified color in at least part of a question-relevant scheme ("Color-coding" and "Color-context" combined), regardless of whether the question prompted about the use of color or a particular scheme.

From the "Color-coding" responses, we see that the *strategic* use of color in instructional materials was considered beneficial for a consistent, non-trivial proportion of students. Across all responses, just under 16% described a relevant color-coding scheme as helpful. In contrast to the total proportion of responses which described *any* aspect of color as helpful, the proportion of "Color-coding" responses did not increase substantially with more specific probing into the use of color. While there was a moderate increase between Question 2 (no specific color prompt) to Question 5 (first to directly ask about color), the proportion of "Color-coding" responses for most questions was within a few percent of the average. The question with the highest proportion of "Color-coding" responses was Question 8 (2D components), which could be attributed to the question text using the word "distinguish". We do not believe this to be the case since all "Color-coding" responses included details beyond those in the question prompt.

Overall, the "Color-coding" results indicate a steady population of students who both consciously noticed color-coding, even though attention was not called to it directly, and were able to clearly articulate its use. Due to the mechanism for anonymizing responses, we cannot know if the "Color-coding" response population was the same for each question, but the steady proportion is suggestive of this possibility.

Responses which described color in a physics context without acknowledging strategic use had a larger variation in response rates, mostly due to Questions 6, 7, and 8 asking about specific contexts while Questions 2 and 5 did not. The proportion of "Color context" responses roughly doubled when prompted to consider the slides in relation to their use of color (Q5) rather than their format generally (Q2), which means that some students needed to have their attention directed to presence of color before they recognized the aid it provided. On the other hand, the overall *ratio* of combined "Color-coding" and "Color-context" responses to "Color generic" responses did not change substantially between these two questions. We infer that when students noticed color, about half articulated it as helpful when used in a physics context, regardless of whether they had been prompted to consider color.

*C. Schemes identified as helpful*

A cursory glance at the "Type" results might imply Scheme 3 to be the most helpful since Question 8, the question which asked about it, had the highest proportion of "Color-coding" responses. Closer examination shows that Schemes 1 and 2 were favored more.

First, within the combined pool of Question 2 and 5 "Color-coding" responses, over half described Scheme 1 (definitions and equations or variables). This could be attributed to greater exposure since Scheme 3 was introduced shortly before Question 5. However, for Question 8, asked in the second-to-last week of the semester, the ratio of *combined* "Color-coding" and "Color context" responses to "Color generic" responses was roughly the same as that for Questions 2 and 5, meaning that among those who found color helpful, the proportion which identified color in context as helpful was similar for Scheme 3 as for any scheme in Questions 2 and 5. Recall also that Question 8 had the lowest rate of positive color responses out of the questions that asked about color. By contrast, for Question 6 and especially Question 7, the proportion of "Color-context" responses was significantly higher, accounting for the high rate of positive color responses for these two questions. A higher proportion of those who found color helpful identifying color in context as helpful for Questions 6 and 7 further supports Scheme 1 (Q6) and especially Scheme 2 (Q7) as being considered more helpful than Scheme 3.

Next, we examined responses classified as either "Color-context" (all questions) or "Other color-context" (Q6-8 only) with respect to what part of a scheme was identified. We flagged keywords related to verbal, mathematical, and visual representations in these responses and tabulated their frequency (Table IV). Many responses included multiple keywords; the sum of sub-categories does not equal 100%. The keywords "component" and "vector" were categorized separately because they each have multiple connotations. "Component" can have the sense of "portion of a whole" (e.g. "visual components" vs "pictorial components" of the slides) or the mathematical sense of "spatial dimension". Similarly, "vector" can refer to a physical quantity conceptually (e.g. velocity or force, in contrast to time), or to a specific representation such as a variable or an arrow in a diagram. In most responses that used these words, the exact sense of the word was not specified.

Looking at the distribution of keywords for "Color context" and "Other color-context" responses, we see that for the more general questions (Q2 and Q5), verbal representations were never referenced and mathematical representations (associated with Schemes 1 and 2) were mentioned more times than visual representations (associated with Scheme 2). Combined with Scheme 1 being the most frequently identified scheme in these questions' "Color-coding" responses, this suggests that Scheme 1 was the simpler color-coding scheme to notice and to describe sufficiently clearly for us to be confident of their noticing.

For Questions 6, 7, and 8, the representations mentioned are consistent with the scheme referred to in the question. More interesting is that aspects of mathematical representations were the most mentioned partial scheme for *all* questions. Although "equations" and "math" were mentioned before "definitions" and "diagrams" in Questions 6 and 7, which could skew responses, the higher proportion across all questions supports this being a real effect. This suggests that while students may not have fully recognized the use of color in connecting equations to other representations, they found the use of color in mathematical representations helpful in and of itself.

**Table IV.** *Percent of responses mentioning specific contexts of color use within the combined pool of Type "Color context" and Type "Other color context" responses, categorized by representation. The number in parenthesis after the column header is the number of responses for that question. Twenty-two (22) of the 56 responses included more than one keyword; percentages for representation categories (bold text) may not equal the sum of individual key word percentages (plain text), and percentages for all representation categories may exceed 100%.*

| Context keywords | All (56) | Q2 (5) | Q5 (8) | Q6 (18) | Q7 (17) | Q8 (8) |
|---|---|---|---|---|---|---|
| **Verbal representation** | **14.3** | **0** | **0** | **38.9** | **0** | **12.5** |
|   Definition | 12.5 | – | – | 33.3 | – | 12.5 |
|   Words | 1.8 | – | – | 5.6 | – | – |
| **Mathematical representation** | **78.6** | **60.0** | **87.5** | **83.3** | **94.1** | **37.5** |
|   Variable | 26.8 | 40.0 | 25.0 | 27.8 | 35.3 | – |
|   Equations | 55.4 | 20.0 | 75.0 | 55.6 | 70.6 | 25.0 |
|   Values | 5.4 | – | 25.0 | – | 5.9 | – |
|   Problem | 7.1 | – | 0 | 11.1 | 5.9 | 12.5 |
| **Visual representation** | **12.5** | **40.0** | **12.5** | **0** | **17.6** | **12.5** |
|   Diagram | 8.9 | 20.0 | – | – | 17.6 | 12.5 |
|   Objects | 3.6 | 20.0 | 12.5 | – | – | – |
| **Other** | | | | | | |
|   Mathematical + "meaning" | 8.9 | – | 12.5 | 5.6 | 17.6 | – |
|   Component | 7.1 | – | – | – | – | 50.0 |
|   Vector | 3.6 | – | – | 11.1 | – | – |

Additionally, several responses referenced the "meaning" of equations or variables without identifying whether the meaning was related to a color-coding scheme. For example, in response to Question 5, "It is helpful because it shows the different parts of the equation and shows what it is", "what it is" could refer to a verbal or visual representation (color-coding) or to the variable's numerical value. Without more detail, those responses were not strong enough to warrant classification as "color-coding". However, the inclusion of "meaning" suggests that some respondents may have inferred a connection.

*C. Reasons for Color's Benefits*

Across all questions, the proportion of responses which described a specific reason for color being helpful ("Specific Reason Subtotal" in Fig 5) was comparable to the proportion which described a generic reason expected from psychology and cognitive science literature ("Generic Reason Subtotal"). The share of responses which described *any* reason for color's benefit increased when the question specifically asked about color (31% for Q2 vs 76% for Q5), and the share of responses which described a *specific* reason increased when prompted to consider a specific context (Q5 vs Q6-8). This is consistent with the "Type" results: some students needed to have the use of color pointed out for them to acknowledge seeing a benefit to it, but a non-trivial proportion of respondents saw a benefit even when not prompted about color. Additionally, for all questions the proportion of responses stating at least one specific reason for color's benefit exceeded the proportion of "Color-coding" responses (43% vs 16% on average), which indicates that more students perceived a benefit aligned with our motivations for employing color-coding than were able to clearly describe the color-coding schemes.

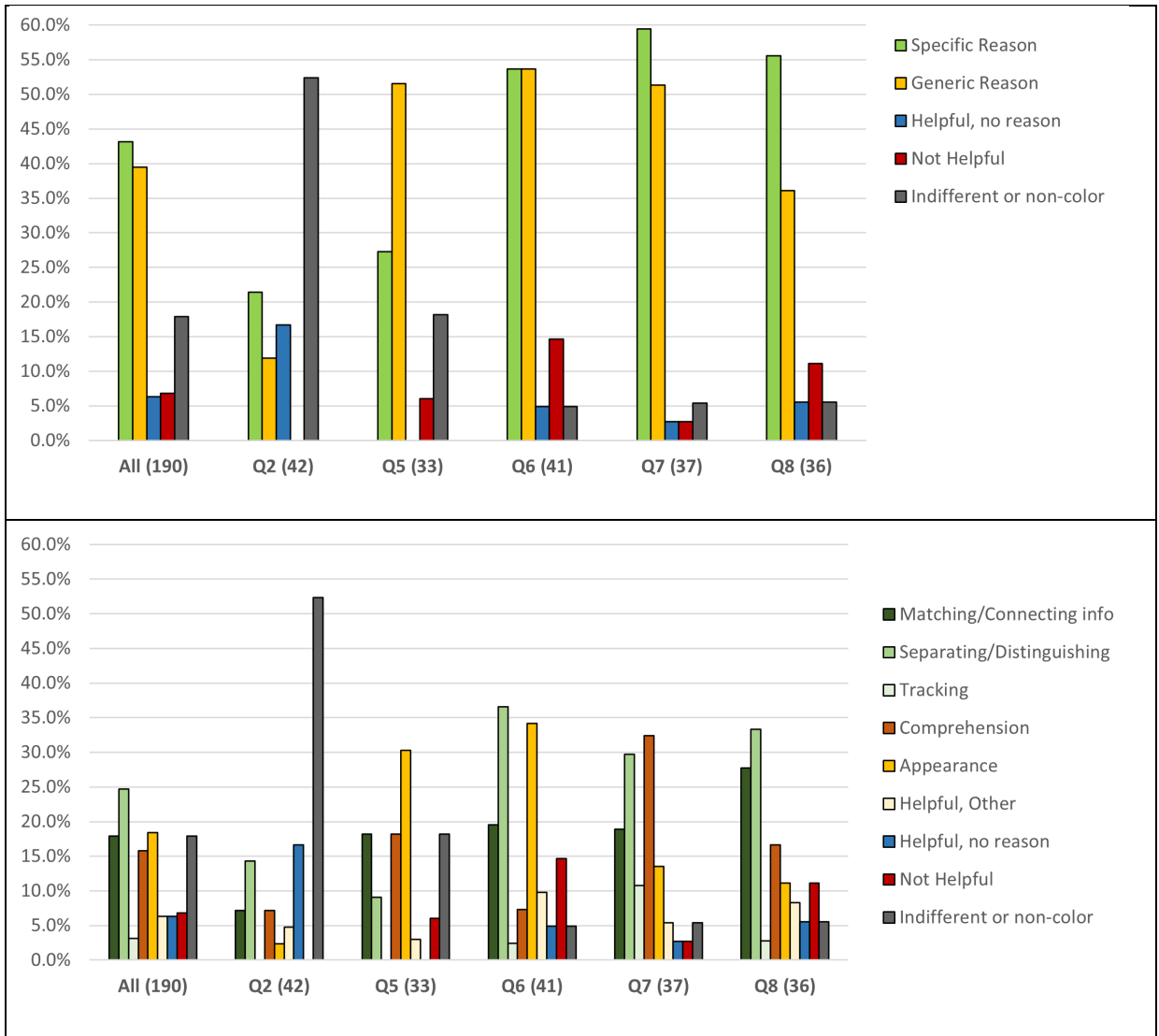

**Figure 5.** (color online) *Percent of responses for each "Reason" category for all responses collectively ("All") and for each question individually, by (top) broad category subtotals and (bottom) individual categories. The number in parenthesis after the cluster name is the number of responses for that question. "Specific Reason" and "Generic Reason" subtotals are the percent of responses which gave one or more reasons within each broad category. Responses may provide more than one reason; sums within a subtotal may exceed the subtotal percentage and the sum of the broad categories for a question may exceed 100%.*

Looking at sub-categories, one immediately notices that the proportion of "Matching and connecting information" responses is very close to the proportion of "Color-coding" responses. This is likely a consequence of our stringent requirements for the "Color-coding" category, with the clearest confirmation that a response recognized color-coding being a description of color used to connect related information. Using color for "Separating and distinguishing information" was the most popular reason on average and the most popular specific reason for all but one question. Tracking information ("Tracking") was found helpful in the context of equations and diagrams (Q7) more than in other contexts; this is understandable since the relevant scheme for Question 7 (Scheme 2) also included problem-solving accompanied by a diagram. Question 7 also had the highest rate of responses which described color as helpful for generic comprehension. Combined with the "Type" results for this question, this suggests that something about the strategic use of color in Scheme 2 contexts was especially helpful to students even if they could not articulate what exactly they found helpful.

Subdividing "Type" categories according to "Reason" category, as shown in Fig 6, further supports the possibility that some students notice the strategic use of color more than they can articulate clearly and that Scheme 2 is particularly beneficial.

Among "Color Generic" responses, one might expect only generic reasons, but a modest proportion described color as helpful for "Separating and distinguishing" information. The fact that the combined proportion of generic and "Helpful, no reason" responses decreases for Questions 6-8 compared to Questions 2 and 5 suggests that at least some "Color generic" respondents noticed the context of color use and responded to the question prompt appropriately, but were too vague in what they described, e.g. "It's good for organizing what into its different parts" in response to Question 6.

Among responses which acknowledged a physics context, "Color-coding" responses found color beneficial for matching information more than for differentiating information, while the reverse was true of "Color context" responses. This was true for every question. However, Question 8's substantially higher proportion of "separating/distinguishing" responses among "Color context" responses may be a less reliable result since that question's wording may have biased responses toward that choice. Lastly, we see that Question 7 having the highest rate of "Tracking" responses is entirely attributable to "Color context" responses. This means diagrams were not explicitly considered alongside following a sequence of math steps.

The striking differences between the reasons given by "Color-coding" and "Color context" responses can be interpreted several ways. Taken at face value, they imply that using color-coding to link different representations was the most noticeable color-coding scheme but not the most helpful, and that students did not consciously realize that color-*coding* was what facilitated separating and distinguishing similar types of information.

Another strong possibility is that students found coloring parts of equations to be helpful even without a link to verbal or visual representations, which was not something our survey questions asked about directly. "Color context" responses not only had higher rates of describing color as beneficial for differentiating information and tracking information, but also mentioned mathematical representations with overwhelming frequency compared to verbal and visual representations (see Table IV). It would be worthwhile to investigate using color to differentiate variables when solving problems without color-coded verbal and visual representations. This is also supported by a small study published shortly before our data collection finished which found students had a favorable response to a simple color-coding scheme for example problems that differentiated knowns, extraneous unknowns, and target unknowns in purely symbolic algebraic solutions.[22]

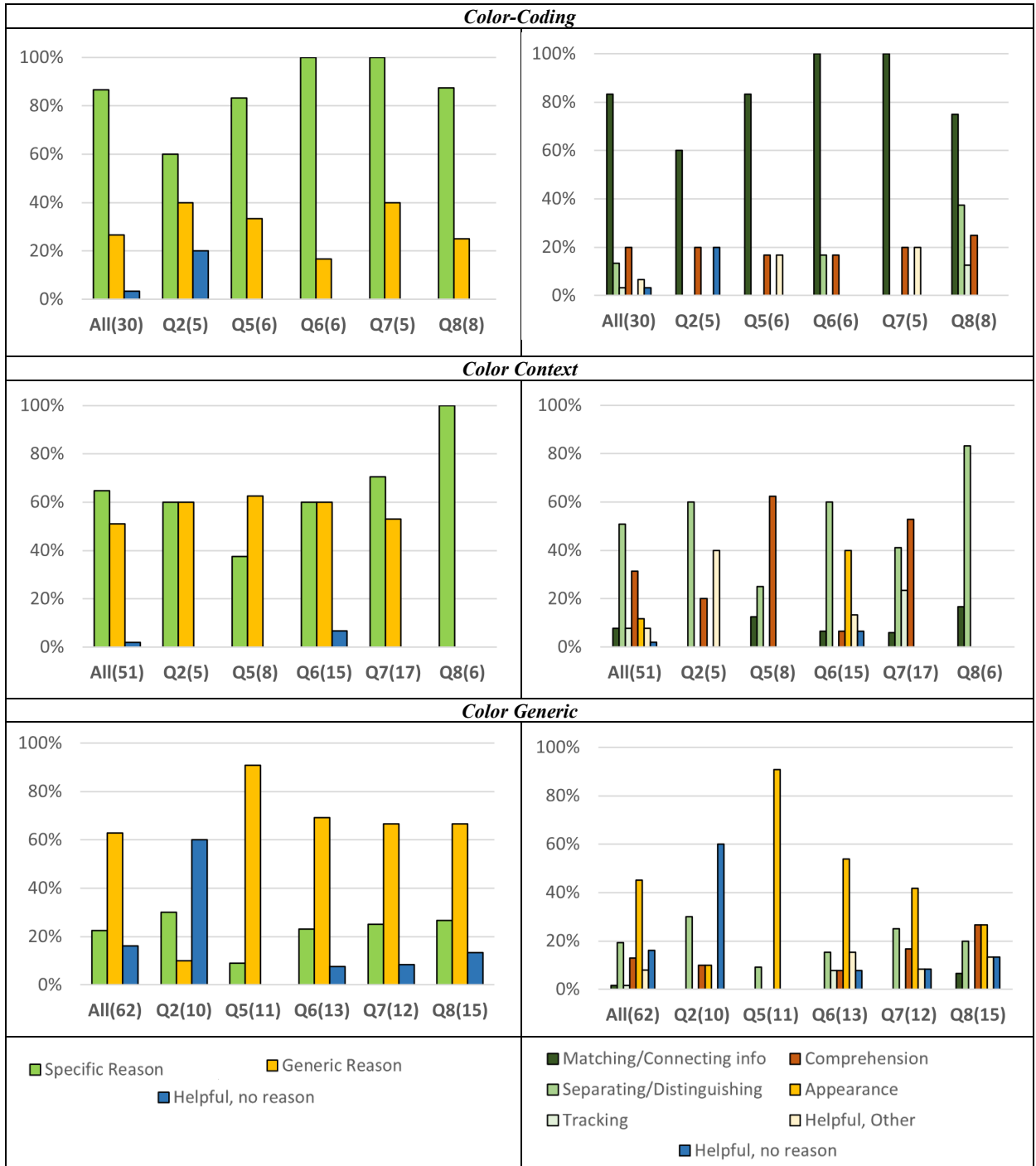

**Figure 6.** (color online) *Percent of responses for each "Reason" category among responses with a "Type" Category for which color is described as helpful, for all responses ("All") and each question, by (left) subtotals and (right) individual categories. The number in parenthesis after the question number is the number of responses the specified "Type" category. "Specific Reason" and "Generic Reason" subtotals are the percent of responses with one or more reasons within each broad category. Responses may give more than one reason; sums within a subtotal may exceed the subtotal percent and the sum of the subtotals for a question may exceed 100%. "Not helpful – reason" is excluded; it is not relevant to these Type categories.*

Lastly, it is entirely possible that strategic color use within a single representation was harder to describe clearly enough to meet the stringent criteria for classification as recognizing color-coding, and that our "Color-coding" percentages are an underestimate.

## V. Conclusion

We draw four conclusions from the results of our study.

First, very few students had a negative opinion on the use of color, and fewer had an exclusively negative opinion. The reasons given for color being unhelpful are easily remedied by simple adjustments to the type and number of colors used and by judicious application of color-coding.

Second, the *strategic* use of color in instructional materials was perceived as beneficial to learning for a consistent, non-trivial proportion of students. The proportion of students who recognized the use of color-coding, were able to describe it clearly, and saw it as helpful to their learning was a minority (on average about 16% of responses), but a consistent one across all survey questions. Furthermore, a little over a quarter of all responses (27%) found color used in specific physics contexts to be helpful even though they did not describe color-coding. Although the proportion of students who identified the use of color or color-coding in a particular context was higher when prompted to consider that specific context, nearly a quarter identified this without prompting about the use of color specifically.

Third, strategic color use for mathematical variables, especially when paired with verbal or visual representations (Schemes 1 or 2), was described as the most helpful implementation, and further investigation should be made on the use of color-coding in mathematics-only contexts. Using color for equations and diagrams (Q7) had the highest rate of responses describing color in any capacity as helpful, the lowest rate of negative or indifferent responses, and no exclusively negative responses. This context also had the highest proportion of students who found the use of color in mathematical representations to be helpful. Using color for equations and definitions (Q6) also had a high proportion of positive responses and was the scheme most frequently described in questions which did not ask about a specific context of color use. Using color for 2D components was most readily recognized as color-coding (Q8), but also had the highest proportion of negative or indifferent responses.

Lastly, the reasons students gave for *why* they considered the use of color to be helpful were aligned with our motivations for employing color-coding: wanting to alleviate student struggles with multiple representations, seeing variables as meaningful, and symbols-only algebra. This was especially true among those whose response mentioned a physics-specific color context, regardless of whether they additionally indicated that the context involved color-coding. On average, as many students found color to be helpful in matching information across representations and in differentiating types of information as found color to be helpful for generic comprehension and aesthetic reasons that were expected from psychology and cognitive science literature.

There are two main limitations to our results. The first is the questions' breadth. Asking what students found to be helpful, not helpful, or indifferent in the same question may have skewed responses toward the first item on the list, helpfulness, more than if there had been a separate response box for each aspect.

The second limitation is the course context. This project was implemented in a flipped classroom with enrollment capped at 42 students per section, both of which are less typical formats for an introductory mechanics course. The slides used for recorded lessons were the same as those which author BSDT used previously for teaching the course in standard format, so the form and style of content delivery was not dramatically different than in a non-flipped classroom. However, the reception of the content was different, since recorded lessons allow for pausing, variable playback speed, and repeating sections. This flexibility may have influenced students' perceptions in some unknown manner. Additionally, the project took place under Covid-19 pandemic protocols, with caps on the number of in-class students and with students able to choose between in-class or synchronous online attendance. The flipped format should have reduced or eliminated the effects of classroom restrictions since all students received the same instruction regardless of attendance modality, but the underlying stress of the pandemic could have influenced student perceptions. On the whole, we believe that the non-standard context of the study implementation does not substantially affect the results, but some care should be taken in generalizing results to larger populations.

Our project's results lead us to conclude that thoughtfully using color for more than just aesthetics is a worthwhile practice. Instructors, and perhaps textbook authors, should consider using color strategically to link different representations and to use color on mathematical expressions rather than restricting color to diagrams and other visual elements. If teaching from a chalkboard or whiteboard rather than slides, color-coding may be more challenging to implement, but could be achieved by underlining related information in color after writing information, or by providing supplemental written materials which use color-coding, such as copies of examples done in class. Although a consistent color-coding scheme should be self-explanatory, drawing explicit attention to the scheme may provide greater benefit, since favorable responses to color increased when drawing attention to its use.

## Supplemental Materials
Data tables for Fig. 4, 5, and 6 are available in online Supplemental Materials.

## Acknowledgements
B. S. Dillon Thomas thanks Eleanor Sayre for several hours of conversation providing guidance prior to designing this project, and her CCU students for their high rate of willingness to participate in this project.

# Appendix
## I. Survey Administration

This study was implemented in a calculus-based introductory mechanics course at a predominantly undergraduate state university (enrollment ~10k students) in the southeastern United States, taught following Eric Mazur's *Principles and Practice of Physics* (1st ed) and using a flipped classroom format with recorded videos of voice-over slide presentations and worked example problems.

Students were notified of the research project on "The effects of the use of color on student perceptions of learning in introductory physics classes" (CCU IRB Protocol #2021.03) in the first week of the semester through a classroom announcement and a digital informed consent document. Survey questions were administered on different days as part of a short daily assignment on the course's online learning management system using an anonymous feedback poll that awarded completion credit. Since polls were anonymous, each survey item had a second required question in which students opted in or out of their response being used for research. This process enabled students to receive credit for their daily assignment regardless of whether they participated in the project.

In addition to not calling attention to color in the instructional videos' narration, other measures were taken to avoid influencing survey responses. The project title in the consent document included only the word "color", not "color-coding", and did not describe the project's research questions. The first and most open-ended survey question was given three weeks after the deadline for opting in or out of the project; the first survey question which referenced color was not until the 6th week of the semester. Additionally, the consent question for each survey item only referenced "[the instructor's] research project" rather than the project title.

## II. Details of Analysis Methodology
### A. Type Categories

The "Type" categories designate both a response's opinion on the use of color and the extent to which a color-coding scheme was described in the response. Detailed category definitions and example responses are in Table A.I.

Each response could have a mix of feedback with different parts categorized as different types. The two non-color categories ("Other non-color" and "Off-topic") and "Color indifferent" have no overlap with the other categories, but a response which stated an opinion on color could describe some combination of relevant color helpfulness ("Color-coding", "Color context", or "Color generic"), non-relevant color helpfulness ("Other color-coding" or "Other color context"), and/or color not being helpful.

Categories were defined with careful attention to responses' precision to avoid reading into responses more than was said. The "Color-coding" category was intentionally the most stringent, requiring either use of the phrase "color-coding", which was never used in any survey question, or a description of all parts of a color-coding scheme in a way that unambiguously acknowledged the use of color in that context as being strategic and purposeful. Similarly, the degree of scheme description required for "Color-coding" and "Color context" responses did not depend on how detailed the question prompt was. This accounted for numerous responses to

**Table A.I.** *Detailed definitions the "Type" categories for coding survey responses, with example responses. Quotes have capitalization errors corrected but spelling and grammar unchanged. The question for which the response was given is in parenthesis.*

| Type Category | Definition | Example Quotes |
| --- | --- | --- |
| **Color-coding** | Identifies color as helpful in a question-relevant color-coding scheme used in class, by either<br>• Explicitly describing the scheme<br>• Describing use of color in a way that indicates respondent recognized pattern matching and/or that color was being used strategically. | *"It also really helps when, within explanations of equations, the color of the variable in the equations correspond with the words that define them in the equation explanations."* (Q5)<br>*"I like the use of colors in the [slides]. It makes it easier to see what pieces of an equation correspond to the diagram or given information."* (Q5) |
| **Other color-coding** | Same criteria as "Color-coding", but identifies scheme not referred to in the question. | … |
| **Color context** | Identifies color as helpful in the context of a question-relevant color-coding scheme but does not clearly acknowledge color-coding or strategic use. May reference all parts of scheme or just one part. | *"I think it is helpful for colors to be used for equations or definitions because then it stands out from the rest of the slides"* (Q6, all parts of Scheme 1, but no connection)<br>*"I think the use of color is really helpful in equations and makes learning them a lot easier."* (Q7, partial scheme) |
| **Other color context** | Same criteria as "Color-context", but identifies context not referred to in question. | … |
| **Color generic** | Identifies color as generically helpful or beneficial, but does not identify color used in the context of one of the three primary color-coding schemes. | *"The colors are great at grabbing attention and making the slides more interesting."* (Q7)<br>*"I find it helpful since it makes things stand out like important words that we will need to know and will most likely use in our exercises that day."* (Q7) |
| **Color indifferent** | Identifies color in response, but neutral opinion on use. | *"Color, font type and size generally don't effect [sic] me as long as its [sic] readable."* (Q2) |
| **Color not helpful** | Identifies use of color as being not helpful, regardless of context. | *"What is not helpful is that [use of color] can become overwhelming at some points."* (Q8) |
| **Other non-color** | Does not mention color at all in description of what is/is not helpful. | *"I find the slides to be very useful. There is just enough information to understand, but not too much to not be able to find important detail."* (Q2) |
| **Off-topic** | Did not address any aspect of the question. | … |

Questions 6, 7, and 8 which began with "it" and did not explicitly state which part(s) of the context mentioned in the question were helpful (see the second example quote for "Color generic"), which made it impossible to determine if the writer intended "it" to refer to the scheme elements stated in the question or if they misread the question. Consequently, even for the questions which asked about a specific scheme context, responses could be classified as "Color generic".

*B. Reason Categories*

Responses which provided a reason for why color was helpful or not were tagged with key words and phrases and grouped by similar topics by first author BSDT to determine the "Reason" categories. These are defined with example responses in Table A.II. Multiple "Reason" categories could apply to a single response if it made distinct statements that were not linked to each other. Each distinct statement was assigned a single category.

Although "Reason" categories were defined by looking at keywords, assigning categories to a response required interpreting the writer's intent. For instance, "The use of color is good for distinguishing what the letters in the math problem stand for when we are first learning what it means" (Q2) uses the word "distinguish" but the full context shows that "distinguish" is used in the sense of "identify" not "differentiate". Thus, this response describes

**Table A.II.** *Definitions of the "Reason" categories for coding survey responses, with example responses for each. Quotes have capitalization errors corrected but spelling and grammar unchanged. The question for which the response was given is in parenthesis.*

| Reason Category | Description | Example Quotes |
| --- | --- | --- |
| *Specific Reasons* | | |
| **Matching/connecting info** | Identifies either matching different representations of same physical entity (math/equation, visual/diagram, verbal/written), or connecting parts of different representations as being related. | *"It [the use of color when presenting equations and definitions] is helpful because I can easily match up what each variable in the equation means and how its [sic] used."* (Q2) |
| **Separating/distinguishing** | Identifies being able to see different types/kinds/parts of information as being distinct from each other. Information may be visual, mathematical, verbal. | *"I appreciate the visuals and colors used to help distinguish between variables"* (Q2) <br> *"I like the colors because it helps differentiate between various things"* (Q7) |
| **Tracking** | Identifies being able to follow a piece of information through/across multiple sequential steps within the same representation. | *"I found the color helpful when the equations get algebraically manipulated so I can follow where it goes."* (Q7) |
| *Generic Reasons* | | |
| **Comprehension** | Identifies color as relates to learning generically, e.g. comprehension, understanding, visualizing, memory, and organizing information in the sense of "seeing related parts". | *"The colors make it easier to understand and remember math equations and make it easier to understand diagrams."* (Q7) |
| **Appearance** | Identifies color as relates to aesthetics, e.g. attention, interest, emphasis, importance, organization in the sense of "outlining/headers". | *"It makes all of the important information stand out and stick in my mind."* (Q7) <br> *"It was helpful to keep the slides more vibrant and interesting"* (Q7) |
| **Helpful, Other** | Describes color being helpful/positive in vague general terms only. | *"Yes. It made explanations more distinguishable."* (Q6, full response too short to know what is meant by "distinguishable") |
| **Helpful, No Reason** | Does not indicate a reason for helpfulness. | ... |
| **Not helpful, reason** | Indicates a reason for why color not liked or considered not helpful. Reasons noted but not coded separately. | See discussion in Section V.A |

matching mathematical and verbal representations and belongs with "Matching/connecting info". Similarly, many responses used the word "track" in the sense of comprehension when holding multiple things in mind, not following sequential steps, e.g. "I find the use of diagrams helpful and the colors associated with them because they help me keep track of more ideas in my head in a clearer way." (Q2)

*C. Survey Response Coding*

The authors trained on a 10% sample of survey responses selected by first author BSDT and category definitions were refined through discussion until rate of disagreement was minimized and agreed to be attributable to responses' variable clarity. All three authors then independently coded all 190 pertinent survey responses. When aggregating results, "Type" and "Reason" categories were assigned based on consensus (at least two of the three agree on appropriateness of category).

For all but two responses, raters had unanimous agreement on which survey responses were on-topic and identified color use. Those two were flagged by at least one rater as potentially responding to a complete misreading of the question and were discarded as "uncategorizable."

For the 162 responses assigned to one of the color-related "Type" categories, perfect or adjacent-level agreement between pairs of raters averaged 91%, where adjacent agreement excludes crossing question-relevance boundaries (e.g. "Color context" to "Other color context"). Ninety-eight percent (98%) of responses had unanimous or consensus choice(s) of "Type" category. The remaining 2% had complete disagreement on category but agreement on the response describing beneficial color, so were assigned to the "Color Generic" category, accounting for 10% of all responses in that category.

For the 156 responses which stated an opinion on color (no indifference), the combination of perfect (all choices matching) and partial (at least one choice matching) agreement on "Reason" category between pairs of raters averaged 74%. The rate of agreement was skewed lower by one rater choosing multiple reasons more sparingly than the others; perfect and partial agreement between the other two raters was 80%. Despite the lower pairwise agreement, only 4% of responses had zero overlap (complete disagreement) among the "Reason" categories selected by all three raters.

Responses with complete disagreement on "Reason" category were assigned to the "Comprehension" category (in sub-category "Comprehension – unclear" in Supplemental Materials Table S.II) if raters chose some combination of the first four "Reason" categories, or "Helpful, Other" (as sub-category "Helpful, no consensus" in Supplemental Materials Table S.II) if raters chose a combination of any helpfulness categories. If there was consensus on a response stating multiple reasons but disagreement on appropriate categories, the first category was assigned by consensus choices and an additional category by evaluating remaining choices using the rules for complete disagreement. In most of these cases, raters agreed on a "Specific" reason category but differed on whether the second reason stated was specific or generic.